\documentclass[aps,prb,showpacs,floatfix,superscriptaddress,onecolumn,amsmath,amssymb,nobalancelastpage,preprint]{revtex4}
\usepackage{dcolumn,bm,amssymb,textcomp}
\usepackage[]{amsmath}
\usepackage{textcomp}
\usepackage{floatrow}
\usepackage{graphicx}
\usepackage{pslatex}
\usepackage{xspace}
\usepackage{type1cm}

\newcommand{\LiZn}{LiZn$_2$Mo$_3$O$_8$\xspace}
\newcommand{\MoO}{Mo$_3$O$_{13}$\xspace}
\newcommand{\MuSR}{$\mu \textrm{SR}$\xspace}

\begin{document}
\title{Local magnetism and spin correlations in the geometrically frustrated cluster magnet \LiZn}
\author{J. P. Sheckelton} 
\affiliation{Department of Chemistry, The Johns Hopkins University, Baltimore, MD 21218, USA}
\affiliation{Institute for Quantum Matter and Department of Physics and Astronomy, The Johns Hopkins University, Baltimore, MD 21218, USA}
\author{F. R. Foronda}
\affiliation{Department of Physics, Oxford University, Clarendon Laboratory, Parks Road, Oxford, OX1 3PU, United Kingdom}
\author{LiDong Pan}
\affiliation{Institute for Quantum Matter and Department of Physics and Astronomy, The Johns Hopkins University, Baltimore, MD 21218, USA}
\author{C. Moir}
\affiliation{National High Magnetic Field Laboratory, Florida State University, 1800 E. Paul Dirac Drive, Tallahassee, Florida 32310, USA}
\author{R. D. McDonald}
\affiliation{National High Magnetic Field Laboratory, Los Alamos National Laboratory, Los Alamos, NM 87545, USA}
\author{T. Lancaster}
\affiliation{Department of Physics, Durham University, South Road, Durham, DH1 3LE, United Kingdom}
\author{P. J. Baker}
\affiliation{ISIS Facility, STFC Rutherford Appleton Laboratory, Harwell Oxford, Didcot, OX11 0QX, United Kingdom}
\author{N. P. Armitage}
\affiliation{Institute for Quantum Matter and Department of Physics and Astronomy, The Johns Hopkins University, Baltimore, MD 21218, USA}
\author{T. Imai}
\affiliation{Department of Physics and Astronomy, McMaster University, Hamilton, Ontario L8S4M1, Canada}
\affiliation{Canadian Institute for Advanced Research, Toronto, Ontario M5G1Z8, Canada}
\author{S. J. Blundell}
\affiliation{Department of Physics, Oxford University, Clarendon Laboratory, Parks Road, Oxford, OX1 3PU, United Kingdom}
\author{T. M. McQueen}\thanks{mcqueen@jhu.edu}
\affiliation{Department of Chemistry, The Johns Hopkins University, Baltimore, MD 21218, USA}
\affiliation{Institute for Quantum Matter and Department of Physics and Astronomy, The Johns Hopkins University, Baltimore, MD 21218, USA}

\date{\today}

\begin{abstract}
\LiZn has been proposed to contain \textit{S}~=~1/2 \MoO magnetic clusters arranged on a triangular lattice with antiferromagnetic nearest-neighbor interactions. Here, microwave and terahertz electron spin resonance (ESR), $^7$Li nuclear magnetic resonance (NMR), and muon spin rotation (\MuSR) spectroscopies are used to characterize the local magnetic properties of \LiZn. These results show the magnetism in \LiZn arises from a single isotropic \textit{S}~=~1/2 electron per cluster and that there is no static long-range magnetic ordering down to \textit{T}~=~0.07\,K. Further, there is evidence of gapless spin excitations with spin fluctuations slowing down as the temperature is lowered. These data indicate strong spin correlations which, together with previous data, suggest a low-temperature resonating valence-bond state in \LiZn. 
\end{abstract}

\maketitle

\section{INTRODUCTION}

Complex states of matter result from simple interactions which give rise to emergent properties appearing greater than the sum of their parts. There are many examples of such emergent phenomena from macroscopic systems, such as the flocking of birds, to quantum scale systems, such as superconductivity \cite{wu_superconductivity_1987,stockert_magnetically_2011,kamihara_iron-based_2008}, spin liquids \cite{han_fractionalized_2012}, spin ice systems \cite{castelnovo_magnetic_2008}, and heavy fermions \cite{steglich_superconductivity_1979,matsushita_evidence_2005,neilson_mixed-valence-driven_2012,urano_liv_2o_4_2000,zhang_topological_2009}. In solid-state materials, utilizing the geometry of the crystal lattice to force a degenerate classical magnetic ground state (``geometrically frustrated magnetism (GFM)'' \cite{ramirez_strongly_1994}) is a powerful approach to designing emergent quantum states. It has been proposed that GFM could possibly result in superconductivity via doping the resonating valence-bond state \cite{anderson_resonating_1987,anderson_resonating_1973}, be used as materials for quantum computing \cite{ioffe_possible_2002}, and harbor exciting low-temperature physics near quantum phase transitions \cite{lee_local_2000}.

GFM materials usually have single ions as the magnetic building block on frustrated topologies such as triangular lattices \cite{ye_spontaneous_2006}, kagome lattices \cite{freedman_site_2010,han_synthesis_2011}, or corner-sharing tetrahedra \cite{morris_dirac_2009}. \LiZn has a structure built by stacking two-dimensional triangular layers of \MoO clusters \cite{sheckelton_possible_2012}. Each \MoO cluster, or ``molecule'', has a single unpaired electron and is expected to act as a \textit{S}~=~1/2 unit, similar to organic molecular magnet systems \cite{takahashi_discovery_1991}. Due to the small distance between \MoO clusters via oxo bridges, there are strong inter-cluster superexchange interactions. A recent report \cite{sheckelton_possible_2012} suggested that \LiZn thus represents an ideal geometrically frustrated triangular lattice antiferromagnetic system in which there is the formation of a condensed valence-bond state, where the formation of resonating valence bonds coexists with remnant paramagnetic spins, when cooled below \textit{T}~\texttildelow~100 K.

Here we report microwave and terahertz range electron spin resonance (ESR), $^7$Li solid state nuclear magnetic resonance (NMR), and muon spin rotation (\MuSR) spectroscopies of \LiZn. The results of these local-probe measurements show that each \MoO cluster in fact behaves as a well localized \textit{S}~=~1/2 magnetic unit. Further, NMR and \MuSR measurements show a slowing of spin fluctuations yet no evidence for static magnetic order down to \textit{T}~\texttildelow~0.07\,K and are instead consistent with a gapless spin excitation spectrum expected for a resonating valence-bond state.

\begin{figure}[!t]
\includegraphics[width=5.in]{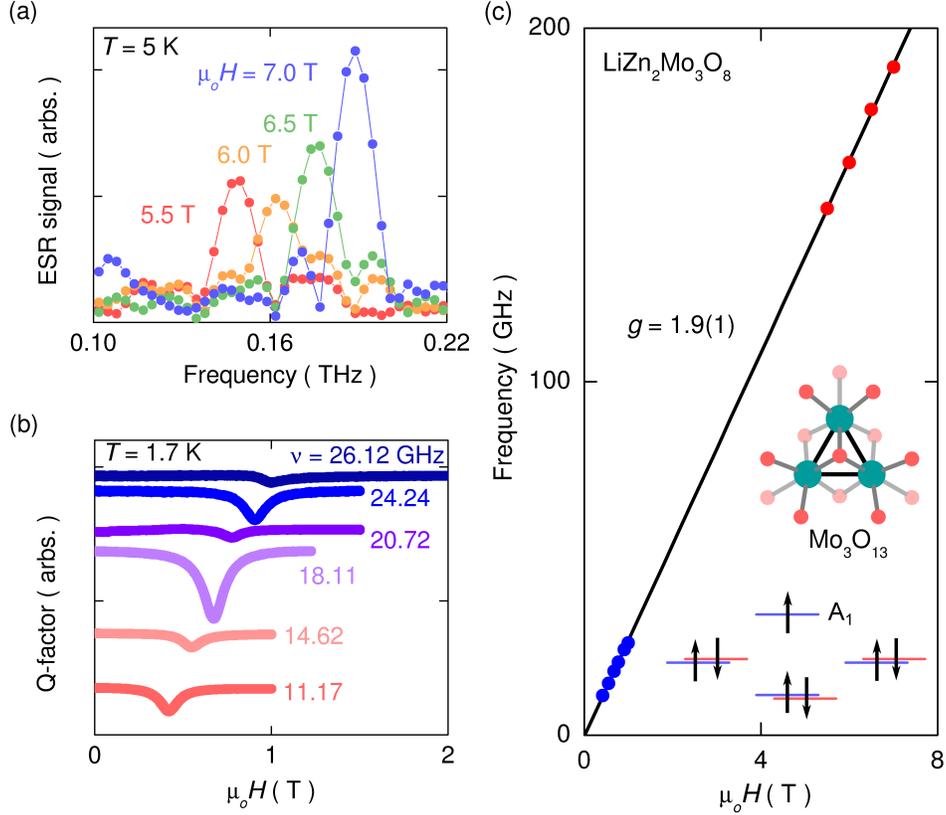}
\caption{ESR spectra of \LiZn at different applied frequencies and fields in (a) the terahertz electromagnetic range and (b) the microwave range (offset for clarity). (c) The extracted \textit{g}-factor for this range of fields and applied EM radiation are in good agreement with each other and are consistent with an isotropic \textit{g}-factor from a single \textit{S}~=~1/2 magnetic electron per formula unit. The magnetic electron is delocalized over a \MoO cluster, shown in the inset with the corresponding molecular orbital diagram \cite{sheckelton_possible_2012}.}
\label{fig:ESR}
\end{figure}

\section{RESULTS}
\subsection*{ESR}

\LiZn powder was synthesized as previously reported \cite{sheckelton_possible_2012}. ESR measurements were performed at the National High Magnetic Field Laboratory (NHMFL) at Los Alamos National Laboratory (LANL) at temperatures from \textit{T}~=~1.7\,K to \textit{T}~=~120\,K in the frequency range of \textit{v}~=~11\,GHz to \textit{v}~=~40\,GHz. Fields up to $\mu_oH$~=~15\,T were investigated, but only one resonance peak was observed, below $\mu_oH$~=~2\,T. The sample microwave absorption was measured using a millimeter wave vector network analyzer manufactured by abmm and a custom-built, appropriately sized resonant cavity. The ESR spectra were measured using cavity perturbation techniques. The signal intensity from the sample cavity of the GHz measurements agree with the number of spins expected for the \texttildelow~1.0 mg sample of \LiZn measured, assuming one unpaired spin per \MoO cluster, and is too large to originate from paramagnetic impurity spins. Terahertz-range ESR was measured at JHU on a home-built transmission-based time domain THz spectrometer using a helium cryostat at temperatures from \textit{T}~=~5\,K to \textit{T}~=~180\,K in the field range $\mu_oH$~=~5.5\,T to $\mu_oH$~=~7\,T using a crossed polarizer geometry \cite{Kozuki11a}.

A single resonance peak is observed, centered at \textit{g}~=~1.9, and is persistent up to \textit{T}~=~80\,K in the THz data with a smoothly decreasing intensity but no change in resonant frequency. The results at \textit{T}~=~5\,K (THz) and \textit{T}~=~1.7\,K (GHz) are shown in Fig.~\ref{fig:ESR}\,(a) and \ref{fig:ESR}\,(b), respectively. The observed \textit{g}-factor, extracted from a fit of the frequency dependence on applied field, Fig.~\ref{fig:ESR}\,(c), is consistent with that expected for a \textit{S}~=~1/2 spin (\textit{g}~=~2.0) and implies minimal orbital contributions. The \textit{y}-intercept of the \textit{g}-factor plot is zero, indicative of a lack of a zero field gap. The combined ESR measurements place a strong constraint on possible \textit{g}-tensor anisotropy \cite{slichter_principles_1990}, assuming additional resonances corresponding to anisotropic \textit{g}-factor values are convoluted within the observed peak, or are outside the frequency/field range probed. By estimating the peak positions and resonance values, the \textit{g}-factors are then constrained to $0.86 \leq g_1/g_2 \leq 1.15$ if the observed peak is a convolution of two resonances or $g_1/g_2 \leq 0.02$ if $g_2 = 1.9$ and an additional resonance lies above the field range measured. The observed peak being a convolution of two peaks is further unlikely as the powder spectrum (angular average) of a system with approximately uniaxial symmetry ($g_{\|}$ and $g_{\perp}$) will have an asymmetric line shape for even small anisotropies, such as in some copper systems \cite{ammeter_static_1979}. These data demonstrate that the magnetism in \LiZn arises from a single, isotropic \textit{S}~=~1/2 magnetic electron per cluster in a totally symmetric (A$_1$) orbital, which is delocalized over the \MoO cluster \cite{sheckelton_possible_2012}. This delocalization over a cluster can represent an intermediate between a tightly bound (\textit{i.e.} fully localized) and completely itinerant electronic wavefunction (\textit{i.e.} a metal) which may contribute to the interesting quantum physics observed in \LiZn. 

\begin{figure}[!b]
\centering
\includegraphics[width=5.in]{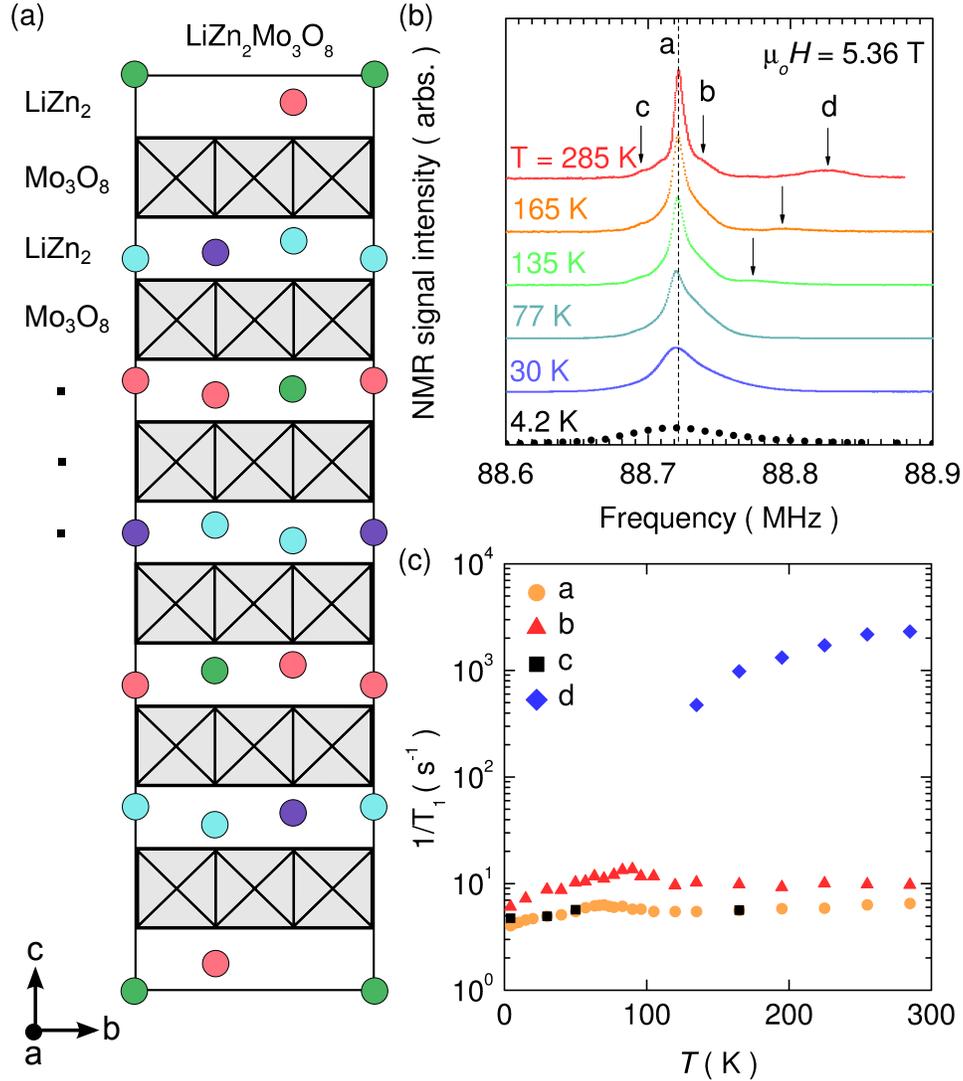}
\caption{(a) A schematic representation of \LiZn crystal structure. Differently colored circles represent four distinct crystallographic, mixed Li/Zn sites. Gray boxes represent Mo$_3$O$_8$ layers. (b) $^7$Li NMR FFT lineshapes taken at $\mu_oH$~=~5.36\,T and point-by-point frequency scan obtained by spin echo integration at \textit{T}~=~4.2\,K (offset for clarity). The spectra indicate four distinct lithium magnetic environments (a-d), consistent with four lithium crystal sites. (c) Spin-lattice relaxation rate as a function of temperature for each NMR peak. The data remain approximately constant as the temperature is lowered, lacking any evidence of the opening of a spin gap.}
\label{fig:NMR}
\end{figure}

\subsection*{NMR}

Cryogenic $^7$Li NMR was measured in the range from \textit{T}~=~285\,K to \textit{T}~=~4.2\,K under an applied field of $\mu_oH$~=~5.36\,T. The NMR lineshapes were measured by taking the fast fourier transform (FFT) of the spin echo. The high frequency (`d' peak) was out of the spectrometer bandwidth, so separately acquired FFT traces were convoluted. Data near \textit{T}~=~4.2\,K were acquired by measuring the integral of the spin echo intensity while scanning frequency. $T_1$ was measured by recording the spin echo integral while scanning the delay time after an inversion pulse. The $^7$Li NMR spectra of \LiZn are shown in Fig.~\ref{fig:NMR}\,(b), displaying at least four peaks (labeled a-d). These peaks may arise from crystographically distinct Li/Zn mixed occupancy interlayer sites (Fig.~\ref{fig:NMR}\,(a)), each being in a unique magnetic environment. Alternatively, the `c' peak could originate from a nuclear spin 3/2 to 1/2 satellite transition, offset from the main peak `a' by a quadrupolar interaction. The `d' peak has a comparatively large frequency shift that merges with the main peak `a' as the temperature of the sample is lowered. The spin-lattice relaxation rate ($(T_1)^{-1}$) as a function of temperature for each peak is shown in Fig.~\ref{fig:NMR}\,(c). The large relaxation rate of the `d' peak is consistent with thermally activated mobile Li atoms ($E_a = 372(26)$~K) at room temperature that freeze out upon cooling, in agreement with previously measured heat capacity data \cite{sheckelton_possible_2012}. The relaxation rate for the `a' site remains approximately constant upon cooling until \textit{T}~\texttildelow~100\,K, then peaks and drops sightly as \textit{T} approaches zero. The low temperature behavior of the relaxation rate indicates the slowing of spin fluctuations and the formation of short range spin correlations. The $(T_1)^{-1}$ measurements agree with bulk magnetic measurements and show that \LiZn does not approach any long-range magnetic ordering, as $(T_1)^{-1}$ would diverge approaching order. Because of the $^7$Li NMR frequency ($\sim$ 88.7 MHz), only the low frequency component of spin fluctuations are detected. The bump in the rate of the `a' and `b' sites is most likely due to the merging of the `d' site intensity and is not intrinsic to those Li sites. As can be seen from the data in Fig.~\ref{fig:NMR}\,(b), the `b' peak is on the higher frequency side of the main peak `a'. The relaxation rate in Fig.~\ref{fig:NMR}\,(c) shows a maxima in peak `b' at \textit{T}~\texttildelow~90\,K, compared to \textit{T}~\texttildelow~70\,K in peak `a'. If this were due to a structural distortion, then the maxima in the relaxation rate for peaks `a' and `b' would occur at the same temperature. Unfortunately, the \LiZn powder $^7$Li NMR FFT lineshapes are too broad (due to powder averaging) to accurately measure the temperature dependence of the knight shift at low temperatures. A close look at the NMR FFT lineshapes, however, reveals a lack of any signature of a split off peak, which would indicate the low temperature bulk susceptibility is dominated by impurity spins, as is the case in other systems \cite{imai_local_2011}. This suggests that the signal from the bulk magnetization measurements is intrinsic to the Mo$_3$O$_8$ magnetic layers in \LiZn.

\begin{figure}[!t]
\centering
\includegraphics[width=5.in]{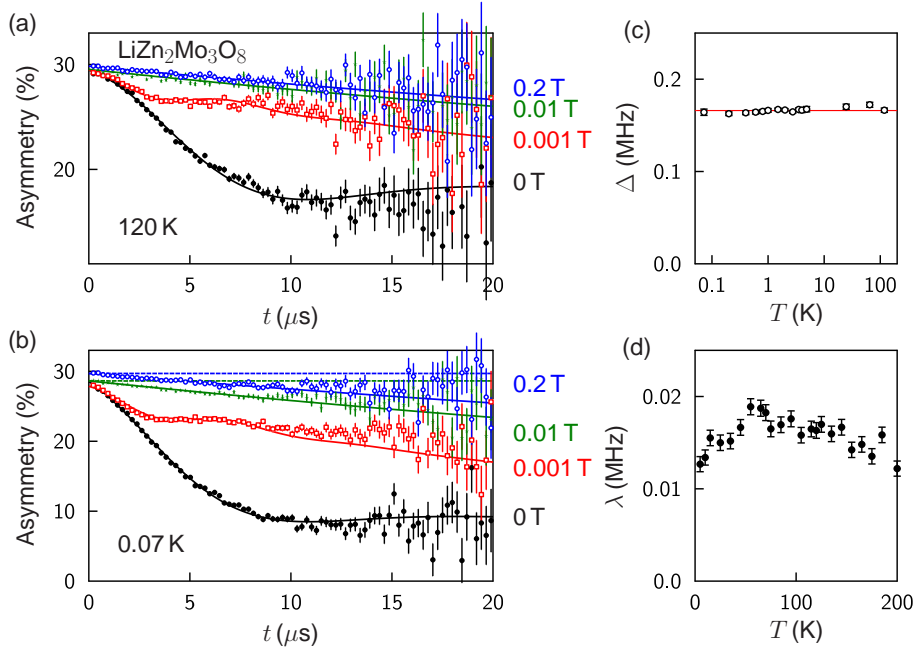}
\caption{Example \MuSR spectra (the asymmetry function $A(t)$) for \LiZn measured at (a) \textit{T}~=~120\,K and (b) \textit{T}~=~0.07\,K in zero field and various longitudinal fields. The fits assume Kubo-Toyabe relaxation with an additional electronic relaxation included (see text). (c) The temperature dependence of $\Delta$, measured in zero field. (d) The temperature dependence of $\lambda$ measured in a longitudinal field of $\mu_oH$~=~10\,mT.}
\label{fig:MuSR}
\end{figure}

\subsection*{\MuSR}

Muon-spin rotation (\MuSR) experiments \cite{blundell_spin-polarized_1999,yaouanc_muon_2010} were carried out using the MuSR spectrometer at the ISIS Pulsed Muon Facility at the Rutherford Appleton Laboratory which is equipped with a dilution refrigerator. In the \MuSR experiment, spin-polarized positive muons ($\mu^+$, momentum 28~MeV$/c$) were implanted into the \LiZn sample. The muons stop quickly (in $<10^{-9}$~s), without significant loss of spin polarization. The observed quantity is then the time evolution of the average muon spin polarization $P_z(t)$, which can be inferred \cite{blundell_spin-polarized_1999,yaouanc_muon_2010} via the asymmetry in the angular distribution of emitted decay positrons, parameterized by an asymmetry function $A(t)$ proportional to $P_z(t)$. The intrinsic low spin-density in this material, due to the cluster-magnet structure, makes it attractive to study using \MuSR at a pulsed source since the internal electronic fields are expected to be relatively small in comparison with other geometrically frustrated materials, particularly those based on rare earth ions. \MuSR is known to be very sensitive to the properties of frustrated systems or those with spin liquid ground states \cite{pratt_magnetic_2011}, as well as magnetic systems with very small ordered moments \cite{wright_gradual_2012,steele_low-moment_2011}. Example \MuSR spectra measured on a sample of \LiZn are presented in Fig.~\ref{fig:MuSR}\,(a) and (b). The zero-field data are characteristic of weak static moments which are nuclear in origin. The data can be fit by a Kubo-Toyabe relaxation function \cite{kubo_magnetic_1967,hayano_zero-and_1979} which models the effect on the muon of a static, random field distribution, multiplied by a slowly-relaxing exponential. The longitudinal Kubo-Toyabe function used in the analysis is given by
\begin{eqnarray}
G_{\textrm {KT}}(t,B_{\textrm {ext}})&=&1-{\frac{2\Delta^2}{\omega^2}}(1-\cos\omega t \, {\textrm e}^{-\Delta^2 t^2/2}) \nonumber \\
&+& {\frac{2\Delta^4}{\omega^4}} \int_0^t
d\tau\,\sin\omega\tau\,{\textrm e}^{-\Delta^2 \tau^2/2},
\end{eqnarray}
where $\omega=\gamma_\mu B_{ext}$ and $B_{ext}$ is the applied longitudinal magnetic field. The data are fit by the function $A(t)=A_1 G_{\textrm{KT}}(t,B_{\textrm {ext}}) {e}^{-\lambda t} + A_0$, where $A_0$ represents muons stopped in the sample holder and $A_1$ is the amplitude of the asymmetry from the sample. This function is parameterized by a variable $\Delta$, the second moment of the field distribution. Fig.~\ref{fig:MuSR}\,(c) shows the fitted $\Delta$ as a function of temperature and it does not change, supporting the hypothesis of nuclear moments. These data thus demonstrate that no static electronic moments freeze out down to \textit{T}~=~0.07\,K.

Application of a longitudinal field partially quenches the relaxation. If the observed behavior were only due to nuclear relaxation, then a field of order $\mu_oH$~=~0.01\,T should be sufficient to completely quench it. The expected behavior if there was only nuclear relaxation is indicated by the dashed lines in Fig.~\ref{fig:MuSR}\,(b). Interestingly, there remains a significant relaxation even in a longitudinal field of $\mu_oH$~=~0.2\,T, the maximum field available on this spectrometer. In fact, the unquenched relaxation observed at $\mu_oH$~=~0.01\,T, which is very similar to that observed at $\mu_oH$~=~0.2\,T, is most likely the result of slowly fluctuating electronic spins, which fluctuate all the way down to \textit{T}~=~0.07\,K.

The temperature dependence of this unquenched relaxation rate $\lambda$ at $\mu_oH$~=~0.01\,T is shown in Fig.~\ref{fig:MuSR}\,(d). Clearly, there is a significant relaxation persisting at all temperatures, indicative of electronic spin fluctuations with dynamics not set by a thermal scale. There is, however, no signature of the valence-bond condensation at \textit{T}~\texttildelow~100\,K that was inferred from bulk magnetic susceptibility and heat capacity measurements \cite{sheckelton_possible_2012}. Rather than indicating the absence of valence-bond condensation, it is possible that the lack of a change at \textit{T}~\texttildelow~100\,K is due to the changes in the spin fluctuation spectrum being outside the muon timescale. Above and below the condensation temperature, the paramagnetic spins fluctuate at the exchange frequency ($J/\hbar \sim 10^{12}\,\textrm{Hz}$), which is too fast to be detected by the muons since muons are sensitive to fluctuations in the \textit{v}~$= 10^5$ to \textit{v}~$= 10^{10}\,\textrm{Hz}$ frequency range. Further, the condensed valence-bonds do not contribute to the local B-field, therefore a transition into a dynamic condensed valence-bond state is effectively invisible to \MuSR spectroscopy. 

\begin{figure}[!t]
\centering
\includegraphics[width=4.in]{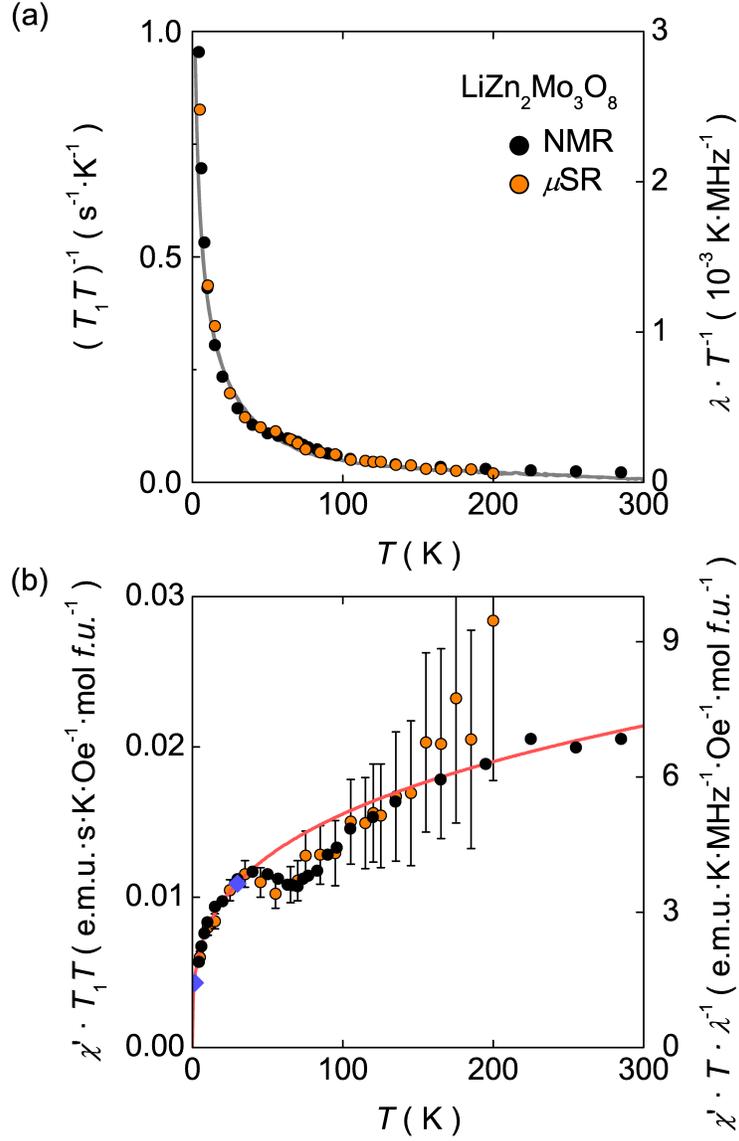}
\caption{(a) NMR spin-lattice relaxation rate (of peak `a', see Fig.~\ref{fig:NMR}\,(b)) divided by temperature, $(T_1T)^{-1}$, a measure of electron spin relaxation scales with \MuSR $\lambda \cdot T^1$. Both datasets are compared to the previously reported bulk magnetic susceptibility, shown as a gray line. The data are self-consistent and indicate gapless, short range spin-spin correlations. The characteristic measurement frequencies for each technique are approximately $\omega_o = 8 \cdot 10^6$\,Hz for \MuSR at $\mu_oH = 10$\,mT and $\omega_o = 9 \cdot 10^7$\,Hz for $^7$Li NMR at $\mu_oH = 5.36$\,T. (b) The bulk susceptibility divided by NMR $(T_1T)^{-1}$ and \MuSR $\lambda \cdot T^1$, a measure of relaxation rate as compared to inelastic neutron scattering data \cite{mourigal_molecular_2013}. The data show a slowing of spin fluctuations as the temperature is lowered, in agreement with the electron spin relaxation rate ($\Gamma$) extracted from inelastic neutron scattering (blue diamonds). The red line is a guide to the eye. The error bars on the \MuSR data were calculated by propagating errors on both the bulk susceptibility and \MuSR datasets.}
\label{fig:NMR-muSR}
\end{figure}

\section{DISCUSSION}

\begin{figure*}[!t]
\includegraphics[width=6.in]{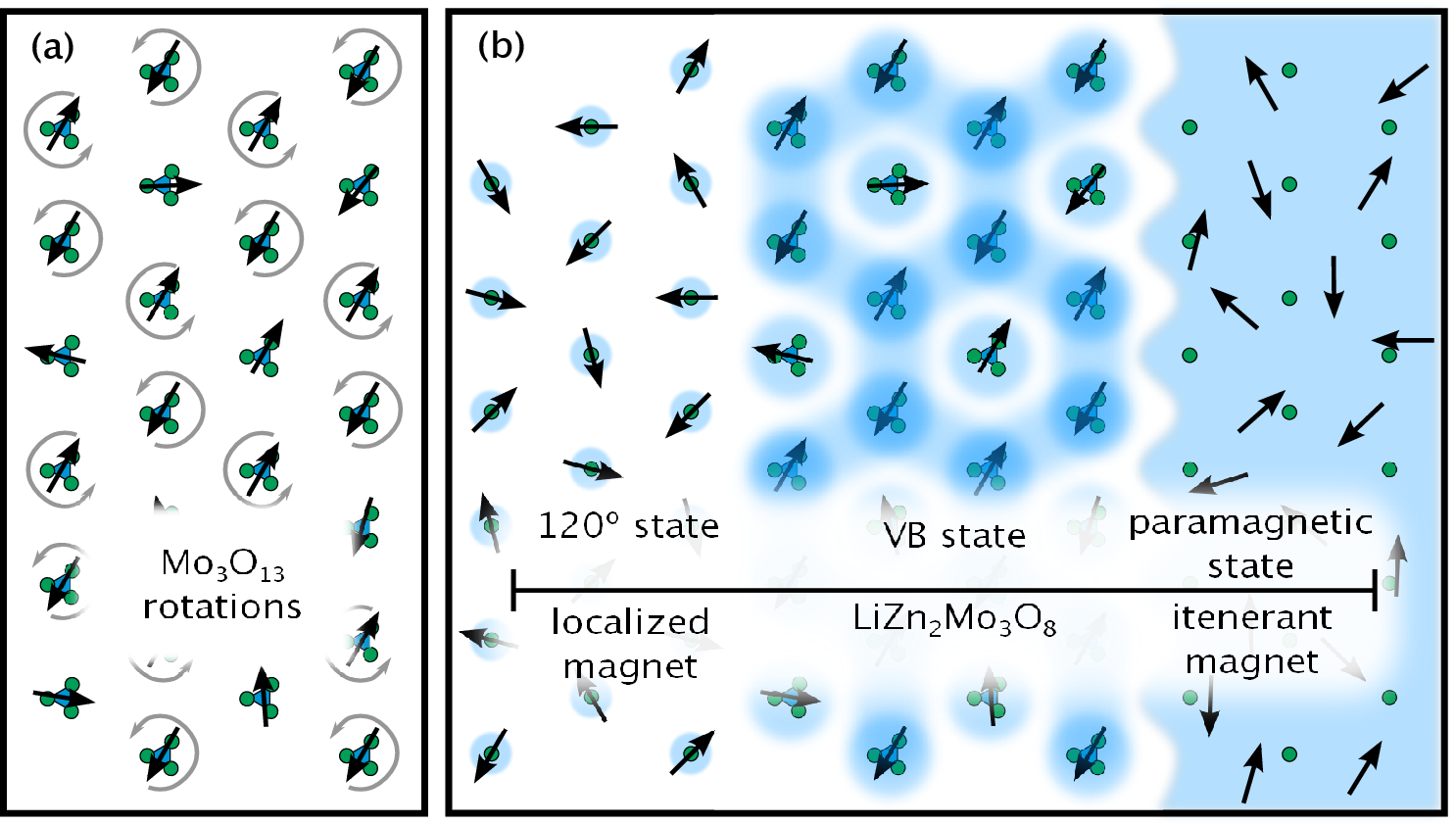}
\caption{Two possible microscopic origins of the valence-bond state in the cluster-based triangular lattice antiferromagnet \LiZn. (a) Dynamic octahedral cluster rotations as proposed in Ref.~\cite{flint_emergent_2013}. (b) Proximity to a metal-insulator-transition. A triangular lattice Heisenberg antiferromagnet has a 120\textdegree\xspace magnetic ground state in the fully localized limit. As the electron wavefunction becomes more delocalized, a valence-bond state, such as what we observe in \LiZn, becomes energetically more favorable \cite{moessner_resonating_2002}.}
\label{fig:phase}
\end{figure*}

More detail about the local susceptibility comes from further analysis of the NMR and \MuSR relaxation dynamics and comparison to bulk magnetization and neutron experiments. The measured \MuSR ($\lambda$) and NMR ($(T_1)^{-1}$, from NMR FFT line `a', Fig.~\ref{fig:NMR}\,(b) and (c)) relaxation rates are related to the local dynamical spin susceptibility since
\begin{equation}
\frac{\lambda}{T} \simeq \frac{1}{T_1T} \simeq \frac{\sum \mid A(\bf q) \mid^2 \chi"(\bf q,\omega_o)}{\omega_o}
\end{equation}
where $A(\bf q)$ is the appropriate wave vector dependent form factor for NMR \cite{shastry_t-j_1989} or \MuSR, $\omega_o$ is the NMR or \MuSR frequency, and $\chi"(\bf q,\omega_o)$ is the imaginary part of the dynamical electron spin susceptibility. Fig.~\ref{fig:NMR-muSR}\,(a) shows $\lambda \cdot T^{-1}$ (\MuSR) and $(T_1T)^{-1}$ (NMR) scaled for comparison to the bulk magnetic susceptibility. These data show, consistent with the bulk susceptibility, that the local magnetism in \LiZn is fluctuating at all accessible temperatures, with an increase in short-range spin-spin correlations, and reduction in the electron spin fluctuation rate, as the temperature is lowered. The discernible deviation or bump of $(T_1T)^{-1}$ and $\lambda \cdot T^{-1}$ in the range $\textrm{50~K} < T < \textrm{100~K}$ is likely due to the freezing of lithium ions. For NMR, this bump originates from the merging of the `d' (mobile lithium) peak with the main NMR peak `a' as the lithium ions freeze. The bump in the \MuSR data is due to changes in $\mu^+$ ion diffusion as the lithium ions freeze, as is well known other systems with mobile lithium ions \cite{ariza_muon_2003,baker_probing_2011,kaiser_li_2000,sugiyama_low-temperature_2010}. More importantly, the trend of increasing $(T_1T)^{-1}$ as \textit{T} approaches zero indicates the onset of short-range spin correlations that do not have a gap in the excitation spectrum. This is in agreement with the dynamical susceptibility extracted from inelastic neutron scattering \cite{mourigal_molecular_2013}, which shows an increase as the temperature is lowered and was suggestive of gapless spin excitations. Fig.~\ref{fig:NMR-muSR}\,(b) shows previously reported \cite{sheckelton_possible_2012} bulk susceptibility divided by the \MuSR ($\lambda \cdot T^{-1}$) and NMR $(T_1T)^{-1}$ which demonstrates the slowing of spin fluctuations as the temperature is lowered. A fit of $\chi"(E)$ to the momentum averaged neutron data yields the relaxation response (from Ref. \cite{mourigal_molecular_2013})
\begin{equation}
\chi"(E)=\frac{\chi' E \Gamma}{E^2 + \Gamma^2}
\end{equation}
where $\Gamma$ is the electron spin relaxation rate and $\chi'$ is the \textbf{Q}-average of the real part of the dynamical susceptibility at $E=0$. The NMR and \MuSR relaxation rate data can be compared to the $\Gamma$ energy scale by 
\begin{equation}
\frac{\lambda}{T} \simeq \frac{1}{T_1T} \simeq \frac{\chi"(E)}{E} = \frac{\chi' \Gamma}{E^2 + \Gamma^2}
\end{equation}
in the limit of $E \rightarrow 0$. This implies that 
\begin{equation}
\frac{\chi'}{(\lambda \cdot T^{-1})} \simeq \frac{\chi'}{(T_1T)^{-1}} = \Gamma
\end{equation}
allowing us to directly compare the \MuSR, NMR, and inelastic neutron scattering relaxation rates, assuming $\chi'(\textbf{Q}) \rightarrow \chi_{DC}(\textbf{Q}=0)$ (\textit{i.e.} no strong form factor effects). As can be seen in Fig.~\ref{fig:NMR-muSR}\,(b) as the temperature approaches zero, the spin fluctuations as probed by all techniques slow down. This indicates that the \MuSR and NMR relaxation rates grow faster than the bulk susceptibility, resulting from an overall slowing of electron spin fluctuations.

Our data demonstrate the gapless, dynamic nature of spin correlations in \LiZn. Previous ab-\textit{inito} density functional theory (DFT) calculations on a Mo$_3$O$_{13}$H$_{15}$ cluster with the same formal electron count as a single \MoO cluster in \LiZn predict the magnetic electron occupies a singly degenerate A$_1$ irreducible (totally symmetric) orbital. This orbital is delocalized over a single \MoO cluster, yet due to \LiZn being an electric insulator \cite{sheckelton_possible_2012}, the electron wavefunction remains confined to the cluster. The strong constraints on the anisotropy of the \textit{g}-tensor for a C$_{3\textit{v}}$ symmetric \MoO cluster imposed by the ESR data are consistent with DFT calculations. The calculations and ESR measurements show that, similar to some radical anion species \cite{ovenall_electron_1961}, the magnetic electron of a \MoO cluster in \LiZn is isotropic in \textit{g}-factor but anisotropic in electron density over the cluster. This suggests that the observed magnetic response of \LiZn does not arise from single ion physics and the lack of any long-range magnetic order originates from a collective interacting \textit{S}~=~1/2 cluster-magnet system. The cluster-delocalized nature of the magnetic electron may also act, among additional interactions, as an energetic destabilization of long-range magnetic order (see Fig.~\ref{fig:phase}).

The antiferromagnetic interactions between cluster spins is mediated by superexchange through Mo-O-Mo oxo-bridges, with the triangular arrangement of clusters in the Mo$_3$O$_8$ layers yielding the frustrated magnetic state. The NMR and \MuSR data show that, as the temperature approaches \textit{T}~=~0, the respective relaxation rates decrease, instead of increasing as would be expected if approaching a long-range ordered magnetic state \cite{ning_^59co_2008}. Comparison to bulk neutron data in Fig.~\ref{fig:NMR-muSR}\,(b) shows that both bulk and local measurements show slowing spin fluctuations as the temperature is lowered. Remarkably, the spin fluctuations slow by a factor of $\sim 4$ from \textit{T}~\texttildelow~300\,K to \textit{T}~\texttildelow~4\,K, which is a low rate of slowing. This is consistent with a spin-frustrated system where the spin-spin correlation length ($\xi$) does not grow much as the temperature is lowered. The spin correlation length is related to the relaxation rate $\Gamma$ by $\Gamma \sim \xi^{-z}$, where $z$ is the dynamical exponent. In \LiZn, the correlation length does not grow much as the temperature is lowered, which is in contrast to other known materials \cite{imai_spin-spin_1993, canals_pyrochlore_1998}. This shows that the low frequency spin dynamics of the electrons in \LiZn is behaving unconventionally and is consistent with the unusual condensed valence-bond magnetic state.

\section{CONCLUSION}

We can speculate why \LiZn does not display a 120\textdegree\xspace magnetic state, as theoretically predicted \cite{huse_simple_1988, bernu_signature_1992, capriotti_long-range_1999} and observed \cite{nakatsuji_spin_2005} for a nearest-neighbor, antiferromagnetically coupled triangular lattice of spins. The simplest classical explanation would be if the spins are not Heisenberg in nature, and thus susceptible to well-known states such as cooperative paramagnetism for Ising spins \cite{gardner_cooperative_1999}; this seems unlikely considering our ESR data. The observed physical properties of a spin liquid state, besides being intrinsic to the spin liquid, can also be attributed to disorder, as conclusively demonstrating the spin liquid state is difficult \cite{han_fractionalized_2012}. The NMR data on \LiZn are suggestive of a minimal defect spin concentration and thus the origin of the magnetic response in this material arises from the Mo$_3$O$_8$ magnetic layers. The lack of magnetic ordering (120\textdegree\xspace state) and the dynamic nature of the spins must arise from more than simple nearest-neighbor, triangular lattice interactions. This does not unequivocally rule out more subtle disorder effects, such as a microscopic distribution of exchange interactions due to Li/Zn mixing between magnetic layers. It is more likely that (static or dynamic) structural distortions predicted to energetically stabilize the condensed valence-bond state \cite{flint_emergent_2013} in \LiZn could contribute to the observed slow spin dynamics. Another realistic possibility in \LiZn is that the more delocalized nature of the spins about the \MoO cluster plays an intricate role, pushing the system closer to the delocalized limit (where the additional degree of freedom and kinetic energy gain of delocalization promotes spin dynamics over spin ordering, see Fig.~\ref{fig:phase}), which is known to result in valence bond states \cite{moessner_resonating_2002}. However, the possibility of longer than nearest neighbor interactions or other effects such as ring exchange \cite{motrunich_variational_2005} causing the dynamic nature of the spins in \LiZn, cannot be ruled out. Regardless of the microscopic details, the data clearly show the dynamic spins in \LiZn are strongly correlated, making \LiZn a spin liquid candidate. Most importantly, our results demonstrate the utility of using cluster-based systems to induce collective magnetic interactions and geometrically frustrated magnetism.

\begin{acknowledgements}
This research was supported by the US Department of Energy, Office of Basic Energy Sciences, Division of Materials Sciences and Engineering under Award DE-FG02-08ER46544 to The Institute for Quantum Matter at JHU. The \MuSR measurements were supported by EPSRC and a beamtime allocation from the Science and Technology Facilities Council (UK). The THz ESR measurements and instrumentation development was funded by the Gordon and Betty Moore Foundation through Grant GBMF2628 to NPA. The work at the National High Magnetic Field Laboratory is supported via NSF/DMR 11574. The NMR work at McMaster was supported by NSERC and CIFAR. JPS would like to thank the William Hooper Grafflin Fellowship. The authors would like to thank W. Hayes, R. Flint, M. Mourigal, P.A. Lee, C. L. Broholm, and O. Tchernyshyov for useful discussions.
\end{acknowledgements}

{\footnotesize
\bibliographystyle{unsrt}
\bibliography{LiZn2Mo3O8_local_mag}
}

\end{document}